# Predicted by Orwell:
## A discourse on the gradual shift in electronic surveillance law.

Scott McLachlan
University of Waikato
Te Piringa
Faculty of Law
scott@mclachlandigital.com

## Abstract

At some point in the history of most nations, one or more events of illegal electronic surveillance by those in power or law enforcement has occurred that has the effect of setting State against Citizen. The media sensationalise these incidents for profit, however they more often correctly express the concern felt by the general public. At these times politicians rise, either into fame or infamy, by proposing new legislation which the public is told will protect them by from future incidents of illegal and unwarranted invasion by officers of the state. Two things have occurred since these protective laws were enacted; technological advancement that is claimed has frustrated legitimate investigation, and changes in the law that are ostensibly presented as intending to facilitate the prosecution or prevention of a publicly decried offence, like child pornography, but which in context deliver expanded powers to the State, effectively weakening the protections previously enacted.

This report looks at human rights legislation in three jurisdictions, starting from a position of comparing and contrasting the protections that are available from illegal search and seizure. By identifying legislative changes related to several forms of electronic surveillance and technology, and the situations that led to them, we can locate the effective peak of protection and discuss the processes that have led to a gradual yet pervasive weakening of those laws in all three nations.

We are regularly diverted by those in power towards disregarding the paranoia of the outliers who have been warning us with their purple prose that big brother is watching. But if we focus on the effect of recent legislative changes in the area of electronic surveillance we can clearly see that the Orwellian dystopia is already here, and we are living it.

# Table of Contents

# Referenced Cases

# Referenced Bills and Acts

# Introduction

The human rights and constitutional laws of many western countries use wording intended to extend an expectation of privacy in our day-to-day lives. This is especially true when we legitimately believe ourselves to be in a private location, such as behind the closed doors of the family home. The expectation of privacy is constantly under fire from a gradual yet pervasive change in surveillance law; laws that have weakened domestic protections and exposed the private lives of citizens, allowing dissemination of information to a wider range of government and non-government organisations. Every change chips away at the individuals' rights, reducing some to the point that they are mere words on paper, no longer supporting the citizens they were enacted to protect.

News articles discussing applications on smart devices that have 'crossed the boundary' are a frequent occurrence. Applications that access and distribute private information stored within the device[1], or worse, accessing the microphone and camera to spy on what we are doing[2,3,4].

---

[1] T Halleck "How to turn off smartphone apps that track you in the background" (2014) <http://goo.gl/zMy6iz>

> Consumer information is being collected and shared by applications on smart devices, often used to target advertising and commercial services. A profit model for developers of some freemium applications has been to integrate third-party background geolocation software code within their app, collecting GPS data and packaging it with browser histories, contact details or other user information and on-selling it to retail and media organisations.

[2] L Freedman "Privacy Tip #7: Who is listening to your conversations through your smartphone microphone?" *Data Privacy and Security Insider*.

[3] M Masnick "Smartphone apps quietly using phone microphones and cameras to gather data *Wireless News* <https://goo.gl/q2rd1Z>

[4] R Schlegel K Zhang X Zhou M Intwala A Kapadia X Wang "Soundcomber: A stealthy and context-aware sound trojan for smartphones" (paper presented at the 18th Annual Network and Distributed System Security Symposium NDSS 2011).

> The authors illuminate on a growing threat from smartphone 'sensory malware'. They demonstrate an embeddable software method that can intelligently, even without a network connection, identify and illicitly capture private information including phone, pin and credit card numbers, entered, dialled on or spoken to the user's smart device. They also present a defensive architecture that could be implemented by device manufacturers to mitigate these attack vectors.

Almost nobody reads the terms and conditions of the device, telephone contract or software they use, clicking through and remaining completely unaware of the illicit ways in which we are tricked into exposing our personal lives[5,6]. In these instances, we freely and unknowingly give up the right to our own information.

Written almost seventy years ago, George Orwell's *1984* paints the bleak picture of a dystopian state continuously watching and collecting information on its citizens via two-way *telescreens*[7]. In contrast to today where smart devices are ubiquitous, not all of Orwell's citizens possessed the expensive telescreens. Today we willingly, and blindly, allow the telecommunications companies, search engine providers, advertising agencies and even the State to have access to the vast array of personal information contained or generated within these devices.

---

[5] S Gindin "Nobody reads your privacy policy or online contract? Lessons learned and questions raised by the FTC's action against Sears (2009) 8 NWJTIP 1.

[6] A Phillips "Think before you click: Ordering a genetic test online" (2015) 11 STL 2.

> Clicking 'agree' to an online contract is deemed as legal consent without any proof that the user read or understood the terms. Counselling the signatory regards the responsibilities and ramifications of agreeing to contract terms should be preferred. Online contracts should be formatted and couched so that the general public can better read and understand, ensuring true 'informed consent'. Online contracts should be flexible, advise who information may be shared with and allow the signatory to opt out or alter the contract terms in a structured manner.

[7] G Orwell *1984* (Signet Classics, London).

# Background and Concepts

While many miss the personal significance, we have all seen numerous stories in the media discussing electronic surveillance. From the stories that came out of the Kim Dotcom case and later discussion of the role of the GCSB in New Zealand, sensational headlines about the American NSA's programs that followed whistle-blower Edward Snowden's publication on the WikiLeaks website, and the more shocking revelations that your local council[8] and hospital[9] are conducting electronic surveillance on a daily basis, over the last couple of years as technology has advanced, the topic has become notorious.

**Commercial Electronic Surveillance**

Commercial surveillance is mostly tolerated; considered by many as another cost of convenience[10], or participation in our consumer-driven society[11,12,13]. Few worry about the

---

[8] Kuringai Council *Overt electronic surveillance in public places* <http://www.kmc.nsw.gov.au/Current_projects_priorities/Initiatives_and_campaigns/Overt_electronic_surveillance_in_public_places>

[9] S Kennedy *Electronic Surveillance in Hospitals: A review* (paper presented at the 4th Australian Information Security Management Conference, Perth, Western Australia, 2006)

[10] *United States v. Jones* 132 S. Ct. 945, 565 U.S.
> Justice Alito at page 10 of her judgement suggests that some people may find the trade-off of privacy for convenience worthwhile, or come to accept this *diminution of privacy* as inevitable.

[11] S Grimes L Shade "Neopian economics of play: Children's cyber pets and online communities as immersive advertising in neopets.com" (2005) 1 IJMP 2.
> The popular NeoPets model known as 'immersive advertising' engages young children through a free website that emotionally invests them to return regularly to care for and entertain their virtual pet; playing games and interacting with the pet and other users in order to earn points. There is a consumerism element to the website where the children are encouraged to spend the points earned on virtual food, clothing, accommodation and toys for their pet – while also being constantly bombarded by real-world fast food and toy advertising that is embedded into every element of the virtual landscape rendered in the game-space.

[12] D Hoffman T Novak M Peralta "Building consumer trust online" (1999) 42 ACM 4.

[13] V Senicar B Jerman-Blazic T Klobucar "Privacy-enhancing technologies: Approaches and development" (2003) 25 CSI 2.

State accessing this information or what it might be used for[14]. Even fewer appreciate how surveillance law has changed since 2011; changes allowing governments and law enforcement agencies greater access to what is an ever-growing body of commercially captured personal information[15].

**Metadata**

The most common and easily accessible form of personal information being collected is classed as *metadata*. Metadata in this context is defined as information *about* a communication, the *Who, when, where* and *how*, but not the *what*, that is; not the conversation that took place during the phone call or the subject line and content of an email message[16]. Metadata is being used by law enforcement and others to draw inferences about our state of mind, our intentions, personal associations and interactions. These inferences can be used to justify stored communications warrants allowing access to the full record stored in ISP and telecommunications providers' servers.

---

[14] L BeVier "Information about individuals in the hands of government: Some reflections on mechanisms for privacy protection (1995) 4 WMBORJ 2.

[15] A Cockfield "Who watches the watchers? A law and technology perspective on government and private sector surveillance" 29:1 QLJ 364 at 7.

[16] Attorney General "Data retention" (2015) <https://www.ag.gov.au/dataretention>
>   This website presents a natural language adaptation of the contents of the Telecommunications (Interception and Access) Amendment (Data Retention) Act 2015, s 187AA(1) for the general public.

While police have been known to mislead judges to secure warrants,[17,18,19,20] they very rarely acknowledge such duplicitous behaviour. Even when caught out[21]. The ability to misinterpret and draw incredible conclusions from the incomplete picture posed by metadata simply provides a new avenue for suitably motivated, well intentioned or even misguided police to achieve warrants they might otherwise be unable to procure in any lawful way.

**Warrants, Scope and New Technologies**

A warrant is presumed necessary before the State may engage in domestic surveillance. Historically warrants had to identify the target individual and mode of communication (specific phone number or internet connection), and therefore there was some certainty that only information pertaining to the target was to be returned. Some recent legislative changes, at least in one jurisdiction, have laid the groundwork for a more scattergun approach, delivering to police the device and subscriber details without the need for a warrant,[22] and recordings of all calls that occurred over a given timeframe or in a given cell site when they have been able to secure one.

---

[17] P Shenon *Secret court says FBI aides misled judges in 75 cases* (2002) <http://www.nytimes.com/2002/08/23/us/secret-court-says-fbi-aides-misled-judges-in-75-cases.html>

[18] A Broyles *Police detective on paid leave after judge alleges he lied to get a search warrant* (2016) <http://kfor.com/2016/02/29/police-detective-on-paid-leave-after-judge-alleges-he-lied-to-get-a-search-warrant/>

[19] J Hopkins *Attorney: Police misled magistrate to get search warrant* (2008) <http://pilotonline.com/news/attorney-police-misled-magistrate-to-get-search-warrant/article_b04fb986-2f3d-5212-8e43-9f7029c3f0e2.html>

[20] C Knaus *Illegal police search causes armed robbery trial to collapse* (2016) <http://www.canberratimes.com.au/act-news/illegal-police-search-causes-armed-robbery-trial-to-collapse-20160318-gnm078.html >

[21] *R v Orry Kuzma* [2016] SCC 150/15
>Two police officers performed an illegal search of an enclosed carport under the pretence of 'seeing if the defendant was home'. During this the female officer observed a television allegedly stolen from a complainant living in an adjacent property. The police officers retreated, and attempted to cover up their illegal search by calling a third officer uninvolved with the case, giving him a false story and asking him to call a magistrate to secure an urgent search warrant. When presented with contradictions in her story of events and other evidence recorded during the District and High court trials of the defendant, Justice Walmsley records that the female police officer maintained the '*ducking and weaving*' and constructively evasive denials she had given in the lower courts. Only when admonished sternly by the Court and repeatedly questioned by defence counsel did she finally admit that she had in fact performed the illegal search, had lied to the third police officer and caused him to mislead a Magistrate into authorising the search warrant used to re-enter the premises.

Law enforcement justify warrantless use of novel electronic surveillance technology by asserting the tech is not considered by existing law, and when privacy concerns are expressed, the fall-back position that the technology has not yet been explicitly ruled out by the courts[23]. Such was the case initially with GPS tracking devices affixed to suspect's cars. It was believed that because the GPS trackers did not record conversations they fell outside of existing surveillance law[24].

This paper begins by providing an overview of the rights afforded to citizens in three jurisdictions; the United States, Australia and New Zealand. This is followed by an overview of the historic legal framework that has come into place to govern electronic surveillance; such situations as police listening to or recording the telephone calls of private citizens. We will look at some of the changes governments have made and are making, whether in shoring up citizens' rights or more often, in simplifying the means by which law enforcement or others can access the ever-growing body of information recorded about individuals. These changes will be constructively illuminated from the perspective of the current technologies that pervade our everyday lives, our altered use patterns, and the potential harm that could come from exposure of our personal information.

---

[22] B Grubb E Partridge *Police scoop up data on thousands in mobile phone 'tower dumps' to track down criminals* (2014) <http://www.smh.com.au/digital-life/mobiles/police-scoop-up-data-on-thousands-in-mobile-phone-tower-dumps-to-track-down-criminals-20140704-zsvtf.html>

[23] A Koppel "Warranting a warrant: Fourth amendment concerns raised by Law Enforcement's warrantless use of GPS and cellular phone tracking" 64 UMLR 3 at 1076.

[24] *Id.* at 1065

Some of the topics that will be discussed include: GPS and newer forms of tracking technology; the systems coming online to give free access to all citizens' internet metadata; law enforcement's use of hacking as an investigation tool, and; how governments are monitoring and firewalling what we see on the internet.

In each of the jurisdictions researched for this report there has been a clear and divisive move towards making more information available to law enforcement agencies with significantly lower thresholds for cause. Certainly, some of the revelations seen in the recent Snowden and WikiLeaks reports would have George Orwell saying; "*I saw that coming!*"

# Constitutions and Human Rights

## Protection against the unwarranted Search and Seizure

This section describes the legislative foundations established in the jurisdictions of the United States, Australia and New Zealand to protect their citizens from illegal search and seizure.

**United States**

The Fourth Amendment to the US Constitution delivers the right of citizens to be secure from unreasonable search and seizure which, by the amendment's definition, are those conducted without a warrant that specifically identifies the place being searched and the person or item sought. While early cases like *Olmstead*[25] focused on fourth amendment rights as protecting property, declining to extend it to the wiretapping of a telephone line, the Supreme Court in *Katz*[26] rejected this property-only approach, extending understanding of the Fourth Amendment and ruling that "…*what a man seeks to preserve as private, even in an area accessible to the public, is constitutionally protected*"[27]. Congress affirmed this position by passing a number of laws in the decade following, in part as a result of the Katz decision but equally in reaction to the wiretapping scandals of the Nixon administration.

Understanding of the Fourth Amendment's position on surveillance was further enhanced by the ruling in *Kyllo*[28], where the Supreme Court ruled that the warrantless use of electronic surveillance technology, in this case a thermal imaging device that accidentally saw heat from lamps used in a garage to grow marijuana, was unconstitutional and violated the reasonable expectation of privacy in ways that would have previously required a warrant and physical search[29].

---

[25] *Olmstead v. United States* [1928] 277 US 438.

[26] *Katz v United States* [1967] 389 US 347.

[27] *Id.* at 350.

[28] *Kyllo v United States* [2001] 533 US 27.

**Australia**

Australia is signatory to the International Convention on Civil and Political Rights (CCPR)[30], yet freedom from unreasonable search and seizure has not been conferred upon its citizens in any direct or meaningful way. The Privacy Act 1988[31] provides the legislative framework for how personal information is to be collected, stored, accessed and shared, however the purpose of this Act is not intended to provide protections similar to the US Fourth Amendment or NZ Bill of Rights Act. An array of statutory provisions and specific powers applies to particular criminal activities, each a defeat to common law of trespass that would otherwise render searches unlawful,[32] such as when NSW Police conducted a raid on hundreds of Greek expatriates in Sydney, searching their homes and arresting family members; they didn't think the effort of getting search warrants was necessary as it was assumed the targets were all non-English speaking migrants who would either be ignorant of or too afraid to invoke their legal rights.[33] The focus of these legislated provisions is to facilitate police investigation[34]. Australian courts approach from a position of strict adherence to the letter of the law, deeming conduct by officers that falls outside the prescribed provisions as unlawful, and any evidence gained, excluded[35]. In some instances, warrantless searches may be undertaken. For example, in some Australian states police have a general power allowing them to search person and vehicle where they reasonably suspect evidence of a criminal offence[36, 37].

---

[29] *Id.* at 40

[30] International Convention on Civil and Political Rights, Art 17.

[31] Privacy Act 1988 at Schedule 1, Australian Privacy Principles (APPs).

[32] P Marcus V Waye "Australia and the United States: Two common criminal justice systems uncommonly at odds" (2004) 224 WMLSSR at 38.

[33] P Grabosky *Wayward Governance: Illegality and its control in the public sector* (Aust. Institute of Criminology, Canberra, 1989)

[34] P Marcus (29) at 38.

**New Zealand**

Like Australia, New Zealand is a signatory to the CCPR[38] who has also not directly incorporated it into legislation. However, New Zealand has developed human rights legislation that echoes many CCPR themes. The New Zealand Bill of Rights Act (BoRA) Section 21[39] provides functionally similar protections to the US Fourth Amendment and is primarily concerned with the privacy interests of individuals against intrusion by the State. Chief Justice Elias referred to this as "*the right to be let alone*"[40]. This right is further enhanced by the Search and Surveillance Act (SSA)[41] 2012, which describes its purpose as "*recognis[ing] the importance of the rights and entitlements affirmed in other enactments including New Zealand Bill of Rights Act 1990.*" [42]

---

[35] *Bunnings v Cross* [1978] 141 CLR 54 at 64.

[36] Police, Powers and Responsibilities Act 2000, ss 27-30.

[37] Summary Offences Act 1953 (Sth Aust), s 68.

[38] International Convention on Civil and Political Rights, *Chapter IV: Human Rights*, Signatories.

[39] New Zealand Bill of Rights Act 1990, s 21.

[40] *Hamed v R* [2011] NZSC 101, [2012] 2 NZLR 305, (2011) 25 CRNZ 326 at [10].

[41] Search and Surveillance Act 2012.

[42] SSA, s 5.

# Legal Frameworks and Recent Changes

## General Issues and Discourse

**United States**

Passing of the Foreign Intelligence Surveillance Act (FISA)[43] in 1978 enabled the creation of Foreign Intelligence Surveillance Courts (FISA Courts). Most used by the NSA and FBI, FISA Courts were originally established to oversee issuing of surveillance warrants against foreign spies' resident on US soil[44].

The government back-peddle on American surveillance law has been traced to the Electronic Communications Privacy Act[45] (ECPA) of 1986[46]. While extending the citizen's ability to sue law enforcement for unlawful breaches to privacy after the fact[47], this law has been interpreted by judges as facilitating warrants for law enforcement to secretly enter private homes and install listening devices, or 'bugs'[48]. Incorporated into the ECPA is the Stored Communications

---

[43] Foreign Intelligence Surveillance Act of 1978, 92 Stat. 1783.

[44] D Cohen J Wells *American national security and civil liberties in an era of terrorism* (New York: Palgrave Macmillan).

[45] Electronic Communications Privacy Act, 100 Stat. 1848.

[46] Get Legal "Electronic Surveillance" (2014) <http://public.getlegal.com/legal-info-center/electronic-surveillance/>

[47] *Id.* at para 6.

[48] *Id.* at para 5.

Act (SCA)[49]. The SCA treats email differently depending on a number of factors. While metadata elements such as the parties involved and time when the email was sent do not require a warrant, the subject line and content in most circumstances do. Table 1 illustrates how email contents are treated by the SCA.

Text within 18 U.S.C. section 2703[50] allows for an administrative subpoena, known as a National Security Letter (NSL), to be served on a records holder. The administrative subpoena requires the records holder to disclose subscriber information including; name, address, telephone and internet connection details, dates and times of access and the details of the service used, IP addresses and the payment methods used by the subscriber[51].

Table 1: US Law Enforcement access to email content[52]

| Email State | Requirement for Access | Statute |
| --- | --- | --- |
| Email in transit | Warrant | 18 U.S.C. § 2516 |
| Email stored on home computer | Warrant | 4th Amendment |
| Email in remote storage, already opened | Subpoena | 18 U.S.C. § 2703 |
| Email in remote storage, unopened and stored for less than 180 days | Warrant | 18 U.S.C. § 2703 |
| Email in remote storage, unopened and stored for more than 180 days | Subpoena | 18 U.S.C. § 2703 |

The ECPA enables disclosure of information to third parties, especially when that third party is an agency of the Government[53]. U.S.C. section 2702[54] describes disclosure as always justified where it is reasonably believed that an emergency is imminent involving immediate danger of death or serious physical injury to any person. This amendment came as part of the USA PATRIOT Act has allowed law enforcement to forgo even the minimum burden of a subpoena or court order when claiming an undefined emergency necessitates access to the records[55]. Although the law makes it voluntary for the service provider to respond, in practice many do[56].

---

[49] Stored Communications Act, 100 Stat 1848, 1860.

[50] USA PATRIOT Act, s 2703.

[51] EPIC "Electronic Communications Privacy Act" (2015) <http://epic.org/privacy/ecpa/> at [18]

[52] Adapted from: EPIC "Electronic Communications Privacy Act" (2015) <http://epic.org/privacy/ecpa/>

The 180-day provision in the 1986 text of section 2703 noted in Table 1[57] is still in force today and draws much criticism. Webmail services such as those of Google and Hotmail did not exist when the original law was drafted and the US Congress describe their approach as one that saw email left on a server beyond 180 days as abandoned property, less worthy of protection than those stored for a shorter period[58,59]. This distinction is no longer relevant [60], as email users now have access to unlimited storage and tend to retain more. The provision is also in conflict with other legislation including the Sarbanes-Oxley Act[61] (SOX), which imposes retention rules on business email that would see it stored in corporate and cloud servers for five years[62], well beyond the ECPA's 180 days.

---

[53] *Id.* at [20]

[54] ECPA, s 2702.

[55] EPIC (52) at [20]

[56] EPIC (52) at [20]

[57] ECPA, s 2703.

[58] R Calo "*Against notice scepticism in privacy (and elsewhere)*" (2011) 1027 NDLR at pp 27.

[59] C Doyle "The USA PATRIOT Act: A legal analysis" (2002) CRS Report for Congress, Congressional Research Service.

[60] Calo (58) at pp 27.

[61] Sarbanes-Oxley Act of 2002, 116 Stat. 745.

[62] *Id.* at s 802(a)(1) and 802(a)(2).

On the 27th of April this year Congress approved[63] the Email Privacy Act of 2016 (EPA), a Bill proposed to amend a number of sections of the SCA. Intended to extend the ruling of the U.S. Court of Appeals in the Sixth Circuit in *Warshak*[64] federally, the EPA strikes out many of the post-180 day storage provisions in the existing SCA[65], replacing them with a blanket requirement for warrant meeting federal *probable cause* requirements[66].

The Protect America Act[67] (PAA), signed by President Bush in 2007, significantly altered FISA. The first change removed the requirement for warrants [68,69]. The more significant change, however, was one of context. The original FISA text required that the party being monitored must be a foreign target and the communications being collected were passing to or from the foreign target and were intercepted with a FISA warrant. The PAA requires neither party to be identified as a foreign target, meaning both parties in an intercepted communication may be domestic citizens[70]. The new requirement sought only that the communication "be about" or "with reference to" a foreign target[71]. As Cardy describes[72];

> If Sally in Toledo were talking to George in Austin about their cousin, Jean, who was on vacation in Germany, the Protect America Act permits intelligence agencies to "collect" this conversation without a warrant. The only thing that the government needs to know before proceeding with the collection is that the

---

[63] US Congress *H.R.699 – Email Privacy Act* (2016) <https://www.congress.gov/bill/114th-congress/house-bill/699/text>

[64] *United States v. Warshak* 631 F.3d 266; 2010 WL 5071766; 2010 U.S. App. LEXIS 25415

> The Court held that law enforcement reliance on the post-180 day provision of section 2703 of the Stored Communications Act in order to obtain warrantless access to emails violated the defendants Fourth Amendment rights.

[65] Email Privacy Act of 2016, s 3(1).

[66] *Id.* At s 3(1)(a).

[67] Protect America Act of 2007, 121 Stat. 552.

[68] E Cardy "*The unconstitutionality of the Protect America Act of 2007*" (2008) 18 BUPILJ 171.

[69] R Posner "*Privacy, surveillance and the law*" (2008) 75 UOCLR 245.

[70] Cardy (64) at pp 189.

[71] Cardy (64) at pp 189.

[72] Cardy (64) at pp 189-190.

> communication is about Jean, whom they reasonably believe to be outside of the United States. Contrary to administration and congressional statements, the statute does not even require the government to suspect the subject of the conversation (Jean, in the example) of terrorist activities or of being a threat to national security.

If many believed section 215 of the Patriot Act which gave agencies like the FBI the ability to obtain library and bookstore receipts of private citizens was already an overreach of government powers[73], the PAA violates the Fourth Amendment further by permitting warrantless domestic electronic surveillance and significantly reducing privacy protections for all American citizens[74].

**Australia**

Australian police have a poor history when it comes to interception of private citizens' telephone calls without a warrant. NSW police during the period from the late 1960's through to the early 1980's designed, built and implemented sophisticated devices solely for the purpose of intercepting and recording telephone calls.[75] Over a sixteen year period, the *Electronics Section*, *Technical Support Group* and in its final incarnation, *Technical Survey Unit* of the NSW Police engaged solely in illegal wiretapping. The illegal acts of these police extended to fraud and misrepresentation, and included funding the purchase of ex-lease Telecom motor vehicles which were registered with the police officers personal details, names and addresses. The officers were supplied with uniforms and equipment designed to match the look and corporate feel of legitimate Telecom employees, all in the express effort to gain access

---

[73] A Mullikin S Rahmon "*The ethical dilemma of the US Government wiretapping*" (2010) 2 IJMIT 4.

[74] Cardy (64) at pp 172.

[75] P Grabosky *Wayward Governance: Illegality and its control in the public sector* (Aust. Institute of Criminology, Canberra, 1989)

to Telecom street pillars, junction boxes or private residences in order to install the unwarranted interception devices.[76] In more than one case police unlawfully captured and listened to the protected conversations between lawyer and client.[77] It was the government's unwillingness to probe too deeply under the veil of secrecy that surrounds policing in Australia that allows these and other unlawful acts to occur for as long as they do.[78]

The Telecommunications (Interception and Access) Act 1979[79] (TIA) requires a warrant in order to intercept telephone calls [80]. Requests for an intercept warrant can only be made where there has been an allegation of a *Serious Offences*[81]; those from a prescribed list[82] of crimes that carry a penalty of life imprisonment or imprisonment for a period greater than seven years. Stored communications warrants only require that the accused face allegation of a *Serious Contravention*[83] of the law which, while including a similar list of crimes[84], reduces the limitations to offences that carry a sentence of as little as three years[85], thus making it far easier for law enforcement to achieve the standard necessary for access to already stored communications.

Recent amendments to the TIA via the Telecommunications (Interception and Access)

---

[76] *Id.*

[77] *Id.*

[78] *Id.*

[79] Telecommunications (Interception and Access) Act 1979

[80] *Id.* at s 6C.

[81] *Id.* at s 46(1)(d).

[82] *Id.* at s 5D.

[83] *Id.* at s 116(1)(d)(i).

[84] *Id.* at s 5E.

[85] Legal and Constitutional Affairs References Committee *Comprehensive Revision of the Telecommunications (Interception and Access) Act 1979* (LCARC 2015).

Amendment (Data Retention) Bill[86] (2015) have considerably changed Australia's privacy landscape, making more personal information available to a wider range of organisations. Approved organisations now have unfettered access to an individual's metadata, which exposes who you communicate with and opens your associations to misinterpretation and inference.

Approved organisations no longer have to meet the original restrictive definition of *law enforcement*. The new text substitutes the term *criminal-law enforcement* and expands the scope to include the Australian Navy, Customs and the Border Protection Service[87]. Also, a new broader definition for the term *enforcement agencies* allows for most federal and local government agencies[88] and a range of private and non-government organisations such as those that operate thoroughbred horse and greyhound racing, the RSPCA, Workplace Safety groups and the Clean Energy regulator[89]. The applicant need only be administering a law rendering them able to impose a pecuniary penalty[90] or be involved in protecting public revenue[91]. The application process remains open indefinitely[92] and new organisations are regularly added.

---

[86] Telecommunications (Interception and Access) Amendment (Data Retention) Act 2015.

[87] *Id.* at s 176A(3B)(a)

> The effect of s 176A(3B)(a) has been to grant access to all federal, state, territory and military police forces, the Royal Navy in its role protecting and policing Australia's coastline, and the Customs and Border Protection Services.

[88] *Id.* at s 110A(1).

[89] Attorney General "Enforcement agency applications: Commonwealth agencies which have applied for ongoing access to telecommunications data" (2015) <https://goo.gl/jQHysu>

[90] TIAADRA (78) at s 176A(3B)(b)

> The wording of this section delivers that organisations which may impose fines such as local councils, the Health Care Complaints Commission, the Taxi Federation, the RSPCA, Workplace Safety bodies and those who have roles overseeing the operation of events focused around organised legal gambling such as greyhound and horse racing agencies but it should be noted, not their overseer, the Office of the Racing Integrity Commissioner.

[91] TIAADRA (78) at s 176A(3B)(c).

[92] TIAADRA (78) at s 110A(2).

TIA changes now require ISPs to retain user electronic communication data for at least two years. The reason for these changes becomes apparent when we consider the distinction between the two warrant types afforded by the TIA and the shift in how we are using technology to communicate. Most people have started to make fewer voice calls, favouring instead the use of SMS, email and social media[93]. These are all data-driven communications generating large amounts of data and metadata that traverse service provider networks and create records in their servers. Unwarranted access to an individual's metadata could be used to facilitate the lower threshold for access to *stored communications* warrants; warrants that would be of little value if ISPs were not also storing user's historic communications. The latest round of TIA changes was necessary to complete a legal structure that has taken almost a decade to create. It renders a situation where the privacy of all citizens can be invaded with ease. Where the slightest inference or pretence drawn from a lone text message or email can be used to justify access to otherwise protected *stored communications*, allowing the 'approved organisation' to abuse the privacy of the individual. Even worse, since the same law restrains the ISP and agency from disclosing that the individual's records were accessed, Australian citizens will simply never know.

Like many other countries[94], Australia operates a government-mandated internet censorship list colloquially known in this case as the Great Australian Firewall.[95] Originally proposed as a system to block child pornography[96] these firewalls have become a tool for governments to prevent citizens accessing websites containing political discourse that the current government finds disagreeable,[97] or which it is claimed contains text of a quality considered to be '*lese majeste*'.[98] The list maintained by the Australian Communications and Media Authority (ACMA) has been shown to contain many websites which were otherwise perfectly legal to view, including online poker and anti-abortion sites, Wikipedia entries, websites about euthanasia and suicide, religious sites, and even the websites of a tour operator and a Queensland dentist.[99] To make the point that the ACMA's list was open to abuse, a Wikileaks anti-censorship activist managed to even get a URL from the Wikileaks site that published the Danish censorship list banned, as well as another that contained political content.[100] Under current rules, any Australian website or ISP which hosts a link allowing access to a banned site can be fined $11,000 per day for every day that citizens have access to the banned material. It is disturbing to think that in doing nothing more sinister than booking a holiday or browsing your dentist's website to book an appointment, you, your ISP and the website that linked you can now be held in violation of telecommunications laws overseen by the ACMA.

---

[93] L Srivastava "Mobile phones and the evolution of social behaviour" (2005) 24(2) *Behaviour and Information Technology*.

[94] China operates the best known internet firewall, however other countries including Denmark, Thailand, Norway and the United Kingdom have all been shown to operate government mandated internet firewalls.

[95] D Pauli *The Pirate Party: How to bypass the Great Australian Firewall* (2014) <http://www.computerworld.com.au/slideshow/342549/pirate-party-how-bypass-great-australian-firewall/>

---

[96] Like Australia's, those of Denmark and Thailand were also claimed to be for the purpose of blocking the international trade in child pornography.

[97] A Moses *Banned hyperlinks could cost you $11,000 a day* (2009) <http://www.smh.com.au/articles/2009/03/17/1237054787635.html>

[98] *Lese Majeste* – the act of criticising a member of the Royal Family. Information has shown that almost all of the websites added to the Denmark and Thai firewall lists were added with the notation of *lese majeste*.

[99] A Moses *Leaked Australian blacklist reveals banned sites* (2009) <http://www.smh.com.au/articles/2009/03/19/1237054961100.html>

[100] Wikileaks *Australia secretly censors Wikileaks press release and Danish internet censorship list* (2009) <https://www.wikileaks.org/wiki/Australia_secretly_censors_Wikileaks_press_release_and_Danish_Internet_censorship_list,_16_Mar_2009>

**New Zealand**

New Zealand law incorporates the concept of "*implied license*", allowing police to physically enter private property and "go to the door of a private premises in order to make enquiry of an occupier for any reasonable purpose" [101]. If the officer catches sight of anything that may be evidence of an offence while acting under this implied license, he may make record of it to use as evidence of what was seen, [102] possibly to secure a search warrant or support a later prosecution.

The Law Commission in its report "Search and Surveillance Powers" described New Zealand's search and surveillance law as outdated, identifying that much of it was written decades ago at a time when information mainly existed in hard copy[103]. They claim existing law barely contemplates electronically stored information, that this inherently limits police powers, confounds investigations and prevents convictions[104]. They implore the government to make it easier for police to utilise technology to intercept electronic communications and track private citizens without their knowledge.

When police have secured a warrant, whether for person or location that the person happens to be in the vicinity of, NZ law allows search of the person, their clothing and anything in their possession[105]. This includes laptop computers, tablets, usb keys and smartphones[106]. The possessor can be compelled to provide assistance in accessing data within the device[107], by providing passwords or encryption keys. The wording of Section 88[108] that allows a warrantless search of anyone who has been arrested, provides sufficient flexibility for police to raise vaguely plausible excuses why they might search electronic devices in that person's possession, for example; claiming the device contains GPS data proving the owner was in a location at the time an alleged crime was committed and for which they have been arrested.

---

[101] *Tararo v R* [2010] NZSC 157, [2012] 1 NZLR 145, (2010) 25 CRNZ 317 at [14].

[102] *Id.* at [14].

[103] Law Commission *Search and Surveillance Powers* (NZLC R97, 2007) at 14.

[104] *Id.* at 5.18.

[105] Search and Surveillance Act 2012, s 125(i).

[106] *Id.* at s 125(l).

New Zealand is a member of the "Five Eyes" international intelligence alliance which has been at the centre of a long-running controversy about mass surveillance of the public[109]. New Zealand's intelligence organisation, the Government Communications Security Bureau (GCSB) was caught conducting domestic electronic surveillance of Kim Dotcom at the request of the FBI[110]. Operations against domestic persons were legislatively forbidden[111]. After eventually admitting the mistake[112], and despite emphatic denials that any similar mistake had ever been made[113], a report into the operations of the GCSB later showed that unlawful surveillance had been conducted on a large number of New Zealand citizens[114].

---

[107] *Id.* at s 130.

[108] *Id.* at s 88.

[109] J Liddicoat "Eyes on New Zealand" (2014) Assoc. for Progressive Communications (APC) and Tech Liberty <http://www.giswatch.org/en/country-report/communications-surveillance/new-zealand>

[110] P Bulbeck "New Zealand Government admits to spying bungle in Kim Dotcom extradition case" (2012) Hollywood Reporter <http://www.hollywoodreporter.com/news/kim-dotcom-new-zealand-government-spying-373774>

[111] Government Communications Security Bureau Act 2003, s 14.

[112] Bullbeck at [7]

[113] S Cowlishaw "Spy boss denies mass surveillance" (2014) Stuff <http://goo.gl/fRONKa>

[114] R Kitteridge "Review of compliance at the Government Communications Security Bureau" (2013) <http://goo.gl/Gk4Guh>

Recent changes to the Government Communications Security Bureau Act (GCSBA) 2003 have made some forms of electronic surveillance of New Zealand citizens legal, altering the meaning of section 14[115] to only prevent such surveillance if it is part of foreign intelligence gathering[116]. Domestic surveillance for the purpose of cyber security[117] or on behalf of the SIS, Police or Military[118] is allowed[119]. No limits are placed on the amount of information being collected and shared with overseas intelligence partners, nor on receiving and using information from them about New Zealanders. Information which it is otherwise unlawful for the GCSB to collect on their own.[120]

Information suggests that through the GCSB, the New Zealand government has successfully completed one or more phases of a project enabling wholesale covert electronic surveillance of all citizens[121]. A 2012 NSA inter-organisational capacity report shows the GCSB were awaiting changes to the GCSBA before bringing the project online[122]. These changes would allow the warrantless mass collection of metadata[123]. The 2015 GCSBA amendment appears to have done exactly that. It has also become public knowledge that organisations like Palantir Technologies, a US firm who develop specialised surveillance intelligence and data analytics software for the NSA and CIA, are working with the New Zealand government[124] to deploy their software systems. Described as a 'Data Fusion Platforms',[125] these platforms provide a 'back-end infrastructure for integrating, managing, and securing data of any kind, from any source, at massive scale.'[126] The implication in this instance has been that the Palantir system is being deployed to manage and integrate the wealth of data sources the NZ Government already maintain, with the huge influx of metadata that is either already coming, or soon will be, from the GCSB.

---

[115] GCSBA, s 14.

[116] A Vance "Demystifying the GCSB Bill: Spies and Lies" (2013) Stuff <http://goo.gl/Y9Ga8s>

[117] GCSBA, s 8A.

[118] GCSBA, s 8C.

[119] Vance at [10].

[120] T Beagle "Does the new GCSB Bill give them power to spy on New Zealanders? (2013) Tech Liberty NZ <http://techliberty.org.nz/gcsb-spying-on-new-zealanders/>

[121] G Greenwald R Gallagher "New Zealand launches mass surveillance project while publicly denying it" (2014) The Intercept <https://theintercept.com/2014/09/15/new-zealand-gcsb-speargun-mass-surveillance/> at [2].

[122] Greenwald & Gallagher at [10].

[123] Greenwald & Gallagher at [6].

[124] S Manning *Has the NSA constructed the perfect PPP? (2014)* <http://thedailyblog.co.nz/2014/09/17/has-the-nsa-constructed-the-perfect-ppp/>

---

[125] <http://www.palantir.com>
[126] *Id.*

# Geolocation Tracking

**United States**

When assessing the legality of warrantless GPS tracking, *Moran*[127] and *Gant*[128] both established that the person travelling on public roads had no reasonable expectation of privacy in his movements from place to place. While an application to the court was protocol for real-time cell site tracking of mobile telephones in the US, in accord with *Moran* and *Gant* courts essentially rubber-stamped the application, allowing tracking without consideration of the legality.[129] When the legislative position changed on GPS tracking devices to one requiring a warrant[130] meeting the standard of probable cause,[131] the situation for cell site tracking remained unchanged. Judge Ornstein was the first to question the legal basis for this practice,[132] ruling that orders allowing real-time cell site information were illegal when the standard for probable cause was not met[133]. The *Ornstein Opinion* also recognised in the legislation the concept of a lesser standard of probable cause for access to historic cell site location data[134]; a principle that has crept into the laws of other countries[135].

---

[127] *United States v. Moran* 349 F. Supp. 2d 425 (N.D.N.Y. 2005) at [467].

[128] *People v. Gant* 9 Misc. 3d. 611 (2005), 802 N.Y.S.2d 839 (N.Y. Co. Ct.) at [847].

[129] Koppell (13) at 1080.

[130] *United States v. Jones* 132 S. Ct. 945, 565 U.S. (2012)
>  Five Supreme Court Justices ruled that warrantless GPS tracking constituted a *search* under the Fourth Amendment and that in installing the GPS device on the motor vehicle police had committed *trespass*. The net effect thereby breaching the defendant's reasonable expectation of privacy, potentially revealing when he had been to the psychiatrist or surgeon or any number of other completely private destinations.

[131] *Re Application of the U.S. for an Order* 396 F.Supp.2d 294 (2005) at [304].
>  In quoting Smith, J., Ornstein identifies that by virtue of 18 U.S.C. s 3177 a warrant is now required for the use of GPS tracking devices. Such a warrant can only be issued once the standard for probable cause has been met.

[132] Koppell (13) at 1081.

[133] *Re Application of the U.S. for an Order* 396 F.Supp.2d 294 (2005) at [327].

**Australia**

We have already discussed that Australian law now allows police, and possibly your ex-wife's private investigator[136], free access to the metadata generated by your cell phone and stored on servers at your chosen service provider. All Australian states now utilise Automated Number Plate Readers (ANPR) affixed to overhead bridge stanchions and gantries on highways and motorways, attached to red light camera equipment at high traffic intersections, and on a large number of mobile police vehicles. For up to ten (10) years police retain information including the locations where your number plate was seen, the time and the direction the vehicle was travelling.[137] Over time, the relational database where this information is stored could be used to pinpoint frequent paths of travel or destinations. CrimTrac, a governmental agency providing services to police forces such as linking and sharing each agency's criminal records, fingerprint and photo identification datasets, have proposed a national system whereby all ANPR data would be forwarded to their servers to be made available without warrant to police nationwide.[138]

By identifying the unique MAC address of one or more Bluetooth or Wi-Fi adapters installed either within the car's entertainment system or the driver or passengers' mobile telephones, the BlipTrack system is able to capture journey and location times, along with route information which it delivers in real-time to traffic monitoring centres[139]. BlipTrack utilises numerous directional antennas within each of the sensor units that can be affixed to a chain of light poles along a highway. The manufacturers claim their system returns substantially accurate positional and speed data[140]. The core interface deployed in traffic monitoring centres displays the distance travelled, speed and location of each vehicle from the time a MAC address is first sighted by the system[141].

---

[134] *Id.*

[135] In Australia for example where there is now a significantly lower threshold for stored communications (historic) versus intercepting communications (real-time).

[136] N Ross *What metadata retention looks like: Prepare to be shocked!* (2015) <http://www.abc.net.au/technology/articles/2015/02/19/4183553.htm>

[137] Parliamentary TravelSafe Committee *Report on the enquiry into Automated Number Plate Recognition technology* (2008) <http://www.cabinet.qld.gov.au/documents/2009/apr/gov%20response%20to%20travelsafe%20report%20no%2051/Attachments/Travelsafe%20R51.pdf>

[138] *Id.* at 15.

Once an investigator knows your number plate and cell phone details, the ANPR data collected across Australia could be merged with metadata containing cell tower location histories and BlipTrack location histories[142] to form fairly complete pictures of an Australian citizen's daily routines, using it either as a way of substantiating that you were in a location at the time that a crime was committed, or to predict where you might be at a given time. All brought together without the need for a single warrant.

**New Zealand**

While an intercept warrant would be required in order to access real-time cell-site tracking of an individual, the same cannot be said of Bluetooth or Wi-Fi MAC address tracking. Both types of tracking can deliver real-time location information achieved through the monitoring of a wireless device in a person's possession, with the only differences being that the tracking equipment must be in closer proximity and able to track the individual without the assistance of a third party, i.e. the telephone service provider. NZ Police have already admitted to the use of MAC addresses, even those captured, supposedly by accident, by Google. Suspect's cell phones are being harvested for MAC addresses which are matched against police records where that device's MAC address has been captured near the locations of other reported offences.[143]

---

[139] BlipTrack "BlipTrack Outdoor Sensor" < http://blipsystems.com/outdoor-sensor/>

[140] *Id.*

[141] *Id.*

[142] BlipTrack Bluetooth and WiFi device MAC address capture and monitoring has been installed across 15 motorways in Melbourne, 7 motorways in Sydney, and over 400 locations in south-east Queensland.

Like their Australian counterparts, the New Zealand government has also begun implementing BlipTrack on our nation's highways.[144]

The NZTA, who operate the BlipTrack system on behalf of the government, claim it does not violate the Privacy Act or constitute warrantless tracking as they believe no personal information is being tracked[145]. They claim that the MAC addresses collected are entirely anonymous[146]. It was pointed out in 2013 that under the NZTA's policies on the system it would be possible for anyone without a warrant, including the police, to access the data with details of a suspect's MAC address and receive both historic and real-time location and speed data[147]. While there is no suggestion that BlipTrack data has been used to date in securing a conviction, police are aware of the system and have already had open access to the vehicle and speed information that the NZTA retain for at least three major New Zealand highways.[148]

---

[143] S Bell *Google data a crime fighting tool* (2010) <http://www.computerworld.co.nz/article/356163/google_data_crime_fighting_tool/>

[144] R Young C Vallyon "Money well spent? The challenge of finding primary data to demonstrate sound infrastructure investment." IPENZ Transport Group Conference, Dunedin < http://goo.gl/yRXiV1>

>   Without notice or consultation with the public, BlipTrack was deployed on light poles along the new Waikato Expressway.

[145] Tech Liberty "NZTA's passive electronic monitoring system." (2013) <http://techliberty.org.nz/tag/tracking/>

[146] *Id.*

[147] BlipTrack "BlipTrack Outdoor Sensor" < http://blipsystems.com/outdoor-sensor/>.

[148] K Reid "Request made under Official Information Act 1982: BlipTrack Systems" <https://fyi.org.nz/request/2109/response/7151/attach/7/OIA%201626%20Alex%20Harris%20SIGNED.doc.pdf>

The argument against warrantless cell site tracking was established by the ACLU's amicus brief in *Davis*[149] as an issue that would potentially allow tracking of people when they had an expectation of privacy, for example; within the home or at the doctors surgery[150]. This author questions how tracking the WiFi or Bluetooth MAC address in real-time with BlipTrack can really be any different than tracking cell phones via the sim card and telecommunications service provider. The only change is the removal of a known third party. There are many hundreds of people who live along or within transmitter range of the highways where systems like BlipTrack are deployed. Does being able to see that they are regularly stationary and 'near' the road violate their privacy in the same way that real-time cell site tracking does?

---

[149] *United States v. Davis* [2015] 12-12928.

[150] *Id.*
> In his concurring judgement, Pryor C.J. discusses that the rapid development of technology should militate in favour of judicial caution. This comment leads easily into the discourse of Jordan C.J., who felt that while the Court chose to limit its decision to the world and technology of the day, holding that a person has no reasonable expectation of privacy in cell site tracking data would have impacts on tomorrow, especially as technology improves the accuracy and availability of such tracking data. He went on to express concerns at the notion of government [law enforcement] being able to conduct 24/7 live tracking without an an appropriate warrant. He too confirmed the position that a person has a reasonable (albeit possibly diminished) expectation of privacy and that a court order should be required in such situations.

## Data Encryption

**United States**

In the aftermath of recent legal battles between Apple and the FBI over access to an encrypted iPhone comes a Bill in the US Senate proposing the Compliance with Court Orders Act of 2016.[151] If enacted, this bill is described as outlawing end-to-end encryption.[152] The Bill would effectively make it illegal for an application or service to encrypt data without some simple means or mechanism available that can return it to a plain text readable format.[153] While being very careful in the wording not to authorise the government to *require* any manufacturer to embed such a mechanism, known as a 'back door' into their software or operating system,[154] it is difficult to see how a manufacturer could comply with the intentions of this Bill without doing just that.[155] A manufacturer back door would not only be open to abuse by the Government and law enforcement, but also to hackers and other people that encryption is implemented to protect us from.[156] Many consider the Bill to be absurd,[157] ludicrous in its

---

[151] Compliance with Court Orders Act of 2016 (CCOA), a Bill presented in the US Senate, no. BAG16460 by Senators Feinstein and Burr.

[152] C Freidersdorph *An attack on privacy from the Senators charged with protecting it* (The Atlantic, 2016) <http://www.theatlantic.com/politics/archive/2016/04/an-attack-on-privacy-from-the-senators-charged-with-protecting-it/478146/>

[153] CCOA, s 2(2) – states that any covered entity that receives a court order shall be responsible for providing data in an intelligible format if such data has been made unintelligible by a feature, product, or service owned, controlled, created, or provided by the covered entity.

[154] CCOA, s 3(3)(b).

[155] J Zdziarski *The dangers of the Burr Encryption Bill* (2016) <http://www.zdziarski.com/blog/?p=6046>

[156] *Id.*

[157] *Id.*

technical illiteracy,[158] damaging to both personal security[159] and the American tech economy.[160]

**Australia**

The Defence Trade Controls Act 2012 (DTCA) makes it unlawful for any person to supply, trade, publish or disseminate technology that is listed in the Defence and Strategic Goods List (DSGL)[161]. The DSGL contains technical specifications for encryption, measured by parameters such as "Key Length" and "Field Size". The DSGL classifies any encryption over 512 bits,[162] this leaves the only encryption left unclassified being so weak that it would be imprudent to use.[163] The DSGL doesn't just classify the encryption method itself, but also covers systems, electronics and equipment used to implement, develop, produce or test it.[164] It has been argued that the broad and imprecise language used in the DSGL effectively outlaws learning about and using many fairly innocuous algorithms, as well as some university training, publications and research in the areas of mathematics and computer science.[165]

---

[158] J Kopstein *Congress' new encryption bill, And it's as bad as experts imagined* (2016) <http://motherboard.vice.com/read/draft-encryption-bill-is-everything-we-feared-security-experts-say>

[159] D Welna *The next encryption battleground: Congress* (2016) <http://www.npr.org/sections/alltechconsidered/2016/04/14/474113249/the-next-encryption-battleground-congress>

[160] See Kopstein (70).

[161] Defence Trade Controls Act 2012, Part 2.

[162] The Defence and Strategic Goods List <http://www.defence.gov.au/deco/DSGL.asp>
[163] D Mathews *Even learning about encryption in Australia will soon be illegal* (2015) <http://www.lifehacker.com.au/2015/05/even-learning-about-encryption-in-australia-will-soon-be-illegal/>

[164] *Id.*

[165] *Id.*

**New Zealand**

New Zealand has a history of hiding behind extremely limiting and undefined policy requirements in order to block the exportation of software by Orion Healthcare and Cyphercom that contained elements of encryption.[166] The same rules about encryption being a potential tool or weapon that can be used by terrorists which had limited the sale of Orion Healthcare's Electronic Health Record (EHR) software to US customers because it encrypted patient details, has also been used by MFAT to prevent the export of completely sealed and unaltered SSH and Python application CDs back to the Finnish and Dutch developers who produced and packaged it.[167] Peter Gutmann of Auckland University's Computer Science faculty considered that during the late 1990's and into the 2000's New Zealand's situation was one that:

> "…enjoys the dubious distinction of having the strictest crypto export controls on earth, including unheard-of restrictions on books and publication of academic research. MFAT (and the GCSB's) intent is to make export so difficult that any non-worthless encryption software (and books and academic research) will never leave New Zealand…"[168]

A New Zealand Customs discussion paper recently released stated that Customs officers want unrestrictive powers to force people to divulge passwords to their smartphones and computing devices at the border, and without any requirement for having to show cause.[169] As was the case with government-overseen internet filtering, the claim is that these powers are necessary in helping to detect and apprehend those involved with objectionable material.[170] All too often the words '*child pornography*' are waved at the media by politicians hoping to curry favour and distract us from any other potential use or misuse. Such as when they secretly downloaded the contacts lists and other information from the smartphone of the former wife of Michael Quinlan at Auckland Airport as she was returning to New Zealand from overseas and handed it to police who would later prosecute Quinlan for drugs offences.[171] Or when, without cause, they targeted law graduate Sam Blackman and confiscated his smartphone, laptop and all other electronic devices in his possession.[172] On returning the devices, Customs claimed that it was not Mr Blackman's attendance at a conference in London on Mass Surveillance held by Edward Snowden that prompted the action, but rather it was because of a website accessed from a shared internet connection in a student flat six years earlier, in 2007, that had prompted the action, in spite of the fact that Mr Blackman had come and gone from New Zealand on many occasions between 2007 and this particular instance in 2013.[173]

---

[166] P Gutmann *New Zealand crypto policy – Confusion now hath made his masterpiece* (accessed: 6th May 2016) <https://www.cs.auckland.ac.nz/~pgut001/policy/>

[167] *Id.*

[168] *Id.*

[169] T Pullar-Strecker *Customs downplays password plan* (2015) <http://www.stuff.co.nz/technology/digital-living/67449940/customs-downplays-password-plan>

[170] *Id.*

While Customs were unable to compel Mr Blackman to provide passwords or assistance to unlock the devices at that time, it is possible to see that such a rule would be used not so much as an aid in child pornography cases, but rather as an adjunct to other investigative services, assisting police and the government to access the digital lives that many citizens would prefer stayed behind an electronic lock and key. When passwords are now used not just to view the desktop of a device, but are often tied into the device-level encryption[174], rules that require one person to provide a password to a shared device could inadvertently be used to unencrypt the digital files of any other user of that device.

---

[171] E Gay *Switched on Gardener boss sent to jail* (2013) <http://www.nzherald.co.nz/business/news/article.cfm?c_id=3&objectid=10880603>

[172] D Fisher *Backpacker stripped of tech gear at Auckland Airport* (2013) <http://www.nzherald.co.nz/nz/news/article.cfm?c_id=1&objectid=11171475>

[173] D Fischer *Customs returns seized property* (2013) <http://www.nzherald.co.nz/nz/news/article.cfm?c_id=1&objectid=11172303>

[174] FileVault on Apple OS X encrypts the entire hard drive using XTS-AES 128 encryption. Any registered user can enter their password at startup to unlock FileVault and use the device. A forensic backup taken while the computer is unlocked by one user account would result in an unencrypted view of the entire hard drive, allowing investigators to see the files of every user who has access to the computer system.

# Hacking as an Investigation Tool

**United States**

While hacking by members of the general public remains illegal, U.S. law enforcement has an established history of hacking the computers of criminal suspects.[175,176] The U.S. Justice Department recently announced their wish that federal search warrants should allow law enforcement officials to remotely hack the computers of individuals anywhere who are suspected of a variety of crimes.[177] This resulted in a decision of the U.S. Supreme Court to amend Rule 41,[178] which now allows any magistrate with authority in a district where a crime has occurred to authorise a warrant for remote access (hacking) and search of a computer storage device located '*within or outside the district*'. This grants U.S. federal law enforcement, the FBI, the ability to seek a warrant to hack into any computer where they can show probable cause, potentially even one located outside the U.S.[179]

**Australia**

Police in Australia have not denied the routine use of hacking tools to electronically break into and spy on suspect's computing devices.[180,181] The Law Enforcement (Powers and

---

[175] J valentine-Devries *FBI taps hacker tactics to spy on suspects* (2013) <http://www.wsj.com/articles/SB10001424127887323997004578641993388259674>

[176] J Cox *The FBI spent $775k on Hacking Team's spy tools since 2011* <https://www.wired.com/2015/07/fbi-spent-775k-hacking-teams-spy-tools-since-2011/>

[177] M Keys *Breaking: Outrageous Justice Department hacking into anyone's computer* (2015) <https://www.theblot.com/breaking-outrageous-justice-department-hacking-anyones-computer-7717515>

[178] Federal Rules of Criminal Procedure, Rule 41.

[179] D Yadron *Supreme Court grants FBI massive expansion of powers to hack computers* (2016) <https://www.theguardian.com/technology/2016/apr/29/fbi-hacking-computers-warrants-supreme-court-congress>

[180] B Grubb *NSW Police use hacking software to spy on computers and smartphones: Wikileaks data* (2014) <http://www.smh.com.au/it-pro/government-it/nsw-police-use-hacking-software-to-spy-on-computers-and-smartphones-wikileaks-data-20140915-10h530.html>

[181] D Cornwall *NSW Police cyber hacking could be breaking the law, warns civil liberties expert* (2014) <http://www.abc.net.au/worldtoday/content/2014/s4088458.htm>

Responsibilities) Act 2002 (LEPRA) does allow police to apply for covert search warrants[182], which it is believed are being potentially misused as permission to remotely access and search computers using expensive FinFisher hacking tools without the target suspect's knowledge.[183] Lawyers point out that the covert warrants legislation embedded into LEPRA is premises-based[184], intended to allow police to secretly enter and search premises and the computers they find there, and was not drafted with the notion of remotely installing hacking tools that can perform real-time keylogging, take screenshots and potentially transmit audio and video feeds.[185] Victorian Police successfully petitioned for the ability to remotely access, or hack, suspects computers. The ability to do so came to them within the Terrorism (Community Protection) Amendment Bill 2015,[186] which significantly expanded the powers that came with a covert warrant issued under the Terrorism (Community Protection) Act 2003 (TCPA).[187] When issued under this Act, the police must demonstrate a reasonable suspicion that the person is or has been a member of a terrorist organisation, may be involved in planning or committing a terrorist act, or that there has been some form of terrorist activity or the planning and preparation of an activity on the premises where the computer is located.[188] It is not difficult to see, however, that Police could use this law to go after those who many may dislike, but who would not normally or easily fit the definition of a terrorist. One state premier has previously stated he will 'run every last bikie out of the state', and used Federal Government anti-terrorist laws to ban associations of proscribed organisations, ensuring the bikie gangs came under the umbrella definition of proscribed organisation.[189] His efforts in declaring bikie gangs under an anti-terror law were eventually defeated in the High Court of South Australia, but if the TCPA had been enacted in his state during the period, the Police could have declared every gang member as a terrorist in order to hack their computers and install spyware and trojan software to collect evidence.

---

[182] Law Enforcement (Powers and Responsibilities) Act 2002, s 3(1).

[183] See Grubb (166).

[184] LEPRA, s75B.

[185] See Grubb (166).

[186] Terrorism (Community Protection) Bill 2015

[187] Terrorism (Community Protection) Act 2003, s 6.
[188] TCPA, s 6, Covert Search Warrants.
[189] A Shand *How Cops made bikies cool again* (2013) <https://adamsh.wordpress.com/2013/08/24/outlaw-chic-how-cops-made-bikies-cool/>

**New Zealand**

New Zealand police have never acknowledged hacking the computing devices of suspects,[190] however information released by Kapersky Lab's SecureList security news website shows that the New Zealand police possess seven (7) servers running the Hacking Team's Remote Control System (RCS) Command and Collect (C2) hacking tool,[191] almost twice that of our Australian counterparts.

The Hacking Team RCS system runs on any Android, Apple, Windows Mobile or Blackberry device. It can also be remotely deployed from the RCS C2 server to the suspect's PCs or laptop. The remote injection Trojan runs in the background on the suspect's PC, infecting any mobile device that the suspect happens to connect to it. Once the mobile device is infected, be it smart phone or tablet, the device becomes remotely accessible to the law enforcement officer, who can issue commands from the C2 server. These commands can cause the device to take remote photos, open the microphone and transmit the conversations heard to the officer, capture information such as contacts, calendar appointments, notes and emails and capture keystrokes (including passwords and credit card details). The officer can access and take control of the WiFi, Bluetooth, GPS and cellular components, giving him the ability to see your location in real time, capture all transmissions coming to and from your phone, and even prevent you from receiving selected text messages and emails.[192]

---

[190] S Bell *Police neither confirm nor deny using electronic hacking methods* (2011) <http://www.computerworld.co.nz/article/494468/police_neither_confirm_nor_deny_using_electronic_hacking_methods/>

[191] SecureList *Hacking Team 2.0: The story goes mobile* <https://securelist.com/blog/mobile/63693/hackingteam-2-0-the-story-goes-mobile/>

[192] *Id.*

# Conclusion

The United States, Australian and New Zealand governments are all working to give themselves and their law enforcement agencies unrestricted access into the digital lives of citizens. Many of these changes were promoted as simple updating of old, pre-digital era laws that hadn't kept pace; an assertion that seems plausible but which is unsupported when you look closely at the context and outcomes, especially when many enable the unwarranted surveillance of people's private digital lives without probable cause, notification or recourse.

The two countries in this report with constitutional laws aimed at protecting the privacy rights of citizens have both made broad changes to foreign intelligence laws, laws that are meant to allow pre-emptive surveillance and protection from foreign terrorist threats. These changes have largely changed the focus of those laws in ways that give agencies unwarranted access to unlimited metadata and simplifying access to all domestic communications. Australia's changes make vast databases of metadata held by ISPs available on informal request to an array of agencies, some of whom are not law enforcement, and with no reporting provisions that would allow the public to see when their data has been accessed. The same contradictory law provides Australians with a protective mechanism of complaint to the Telecommunications Ombudsman when information is unfairly or improperly accessed, but when it restrains agencies and ISPs from telling you that someone has accessed your metadata, it effectively renders that protection moot.

Victorian Privacy Commissioner Helen Versey's comments best summated the outcomes that will come from these legislative changes when she said;

> "The vague nature of these items may result in the ability for monitoring of any communication across any telecommunications device without any requirement to show necessity or cause for such monitoring prior to obtaining a warrant."[193]

In concluding, George Orwell's story is beginning to look less like science fiction and more like reality. Each government is working towards a future where we won't know what personal information is being captured… or who has accessed it.

---

[193] L Tung *Aust. State not alone in police hacking laws* (ZDNet, 2009) <http://www.zdnet.com/article/aust-state-not-alone-in-police-hacking-laws/>

# Epilogue

When proposed more than 8 months ago, this Directed Study thesis was going to review cyber law cases conducted before the New Zealand bench, looking at the evidentiary standards and (mis)understandings that can blossom from the not uncommon technology illiteracy of many currently sitting judges. Requests were made through two Barristers of good standing and three court Registrars for electronic copies of the transcripts and exhibits for a list of trial cases identified as having a computer crimes element, however, when it became apparent that starting to write this thesis was essential (to its timely completion, of course), only two 'casebooks' had been received. At the time of writing this epilogue, a third has been received and read, though not so much out of necessity or even morbid curiosity, but rather from a simple need to see if the impressions gained from reading the first two were borne out in the third.

I am very sorry to say; *they were*.

In New Zealand we unfortunately have a significant percentage of jurists from the previous generation. Those that grew up without exposure to computers in the home in their formative years. While I am certain many of them would now blindly assert their expertise with computers, perhaps saying that that they were introduced to computers at university or during their career or at large law firms that tended to adopt casefile systems in the late 80's, it is fair to say that if we were able to ask them about it without engendering the kind of negative reaction that makes them feel confronted or threatened, they would be likely to admit that computers largely represent for them a *black box*, that they have little or no understanding of what actually goes on inside, how it works or how the internet and other networked systems are constructed or operate. Their understanding is largely limited to turning it on, logging in and using the standard applications like email, office apps and web browsing that everyone uses.

Thirty years after Apple, and later IBM, made the personal computer something that many homes and schools could easily acquire, with many homes now having more than one, it is distressing to see jurists who believe that IP addresses are like a license or tax file number – something of a one-to-one nature, identifying a particular person. They do not understand that just because an action was logged or traced to a particular IP address, that this in no way allows us to infer that a particular person did something. An IP address equals a device – a modem, a router, a switch or some other piece of technical equipment that can be used by one or many user devices (laptops, tablets and so on), each of which can have one or more human users. Some being the intended human users, and in the case of hackers or your neighbour who accesses your Telecom Wi-Fi router that is still in its default 'open' and insecure state, some not.

While it never got to be the point of this particular thesis, I would like to conclude by challenging this University to engage law students in at least some introductory computer science training. I do not speaking of training in Cyber Law or Computer Crimes, rather, I refer to actual training in the basics of how computers operate, how they are interconnected, how evidence should be gleaned from them and the new ways in which governments, police, and the general public are using them. Computers are going to feature heavily in their future – whether in Court, in practice, or more likely, in defending or prosecuting clients.

Information technology has become pervasive into almost every facet of human existence. Go to a hospital and the nurses and doctors spend a good percentage of their day working with computers. Banking is done by computers. Every home has computers. Cars are becoming a self-driving collection of computers. Every person you see in the street has a computer in their pocket more powerful, we are told, than those that put man on the moon, and functionally capable of being tracked, data mined and used either for or against its owner. The only way to ensure that the judges of the next generation are not as ignorant of how technology really works as those of the current is to incorporate it into their training.

SM

Sydney
5th May, 2016

## Acknowledgements

Whilst this section usually begins by acknowledging the invaluable assistance of some wise and all-knowing mentor, the majority of this report comes from over five months of research and reading, the combined experience of almost eighteen years in IT and twelve months studying the Law, and a significant dose of life lived at the euphemistically entitled *school of hard knocks*.

My thanks go to the law faculty of Waikato University who extended to me the opportunity of studying in their graduate law program, where I have mostly been engaged in papers from their fourth-year rota. I realise that coming from an IT and Information Sciences background with very little foundation in Law may have raised eyebrows, but I hope that my efforts and results to date justified the decision to accept me.

Also, to the people around me whose efforts have kept me upright, who tolerated my struggle while contemporaneously writing both this and the first half of my MPhil thesis. You know who you are. Thank you.

For Danika, Thomas, Liam and James.

SM

Sydney
7$^{th}$ May, 2016